\documentstyle[12pt,epsf,psfig]{article}
\def\J{$J/\psi$}
\def\j{J/\psi}
\def\X{$\chi$}

\def\P{$\psi'$}

\def\A{$A_{\rm cl}$}
\def\a{A_{\rm cl}}
\def\N{$n_{\rm cl}$}
\def\n{n_{\rm cl}}
\def\S{S_{\rm cl}}
\def\s{s_{\rm cl}}

\def\be{\begin{equation}}
\def\ee{\end{equation}}

\def\lsim{\raise0.3ex\hbox{$<$\kern-0.75em\raise-1.1ex\hbox{$\sim$}}}
\def\gsim{\raise0.3ex\hbox{$>$\kern-0.75em\raise-1.1ex\hbox{$\sim$}}}

\def\NP{{ Nucl.\ Phys.\ }}
\def\PL{{ Phys.\ Lett.\ }}
\def\PR{{ Phys.\ Rev.\ }}

\def\PRL{{ Phys.\ Rev.\ Lett.\ }}

\def\ZP{{ Z.\ Phys.\ }}

\begin{document}

\noindent May 1998~\hfill BI-TP 98/10

\vskip 1.5 cm

\centerline{\Large{\bf String Clustering and J/$\psi$ Suppression}}

\medskip

\centerline{\Large{\bf in Nuclear Collisions}}

\vskip 1.0cm

\centerline{\bf M.\ Nardi and H.\ Satz}

\bigskip

\centerline{Fakult\"at f\"ur Physik, Universit\"at Bielefeld}

\par

\centerline{D-33501 Bielefeld, Germany}

\vskip 1.0cm

\centerline{\bf Abstract:}

We study the clustering of colour strings in the transverse plane of
nucleus-nucleus collisions and argue that deconfinement sets in when
the string density within a cluster reaches a critical value. We show
that this implies a minimal cluster size at the onset of deconfinement,
which in turn results in discontinuous behaviour for \J~suppression.

\bigskip

\vskip 1.0cm

Statistical quantum chromodynamics predicts that strongly interacting
matter will become deconfined at high temperatures and/or densities.
The aim of high energy nuclear collisions is to produce and study this
new state of matter, the quark-gluon plasma, in the laboratory. One
prerequisite for such studies is an understanding of the onset of
deconfinement. In the most general sense, the deconfinement of quarks
and gluons means that these partons are no longer constrained to the
distributions by which they are governed within hadrons \cite{KS-Hwa}.
Such constraints are presumably removed once enough nucleon-nucleon
interactions overlap in space and time, i.e., for a sufficiently high
collision density. A partitioning of the coloured partons into
colour-neutral subsets is then no longer meaningful; instead, there
emerge colour-conducting clusters of much larger than hadronic size.

\medskip

Cluster formation is the central topic of percolation theory, and hence
deconfinement and percolation appear to be closely related \cite{Baym}
- \cite{Pajares}. The aim of the present note is to pursue this
relation in more detail, in particular for the finite-size environment
encountered in nucleus-nucleus interactions, and to apply the results to
recent data on \J~suppression in $Pb-Pb$ interactions \cite{NA50}.

\medskip

Percolation theory \cite{Stauffer} in its conventional continuum form
\cite{Isi} investigates the formation of clusters of differents sizes,
if $N$ extended objects (`discs' or `spheres' in two or three space
dimensions, respectively) are randomly distributed in a given spatial
region, allowing overlap. For reasons to be explained shortly, we
consider here the two-dimensional case, taking the spatial region to be
a circle $A=\pi R^2$ of radius $R$ and the discs as circles $a =\pi r
^2$ of radii $r$ (see Fig.\ 1). The basic question is how the average
geometric size \A~of a cluster of connected discs varies with the
overall disc density $N/A$; it is convenient here to use the
dimensionless density $n=a(N/A)$. The percolation point $n_p$ is defined
to be that density $n$ at which in the limit $R \to \infty$ infinite
clusters appear for the first (when $n \leq n_p$) or last time (when $n
\geq n_p$); in two dimensions, it is found to be $n_p=1.175$
\cite{Alon}, in accord with numerical results \cite{Isi}. Approaching
$n_p$ from below, the average cluster size thus diverges,
\be
\a(n) \sim \left(1 - {n\over n_p}\right)^{-\tau_1}, ~~n < n_p,
\label{1}
\ee
while coming from above, the ratio of average cluster size to overall
area vanishes, with
\be
{\a(n) \over A} \sim \left( 1 - {n_p\over n} \right)^{\tau_2}, ~~n >
n_p; \label{2}
\ee
$\tau_1$ and $\tau_2$ denote the relevant critical exponents.
In all previous work on percolation in strong interaction thermodynamics
\cite{Baym}-\cite{Pajares}, it was assumed that deconfinement sets in at
the percolation point of some specific extended objects. In general,
this conjecture does not seem tenable; in the Ising model, for
example, random percolation and thermal critical behaviour (spontaneous
magnetization) do not coincide \cite{Stauffer,Coniglio}, and pure
$SU(N)$ gauge theory is in the same universality class as $Z_N$ spin
systems \cite{S-Y}. The reason for the difference lies in the
different concept of cluster in the two cases, which tends to force
thermal critical behaviour to higher densities than required for
percolation. Hence a redefinition of percolation clusters
is needed to make the two phenomena coincide; on the lattice,
suitably defined site-bond clusters are shown to do this
\cite{Coniglio}. The generalization of such considerations to the
percolation of extended objects is not clear, however, so that
deconfinement could presumably set in at $n_p$ or at some higher
density; for further discussion, see \cite{satz}. Here we therefore
leave the value of the deconfinement threshold $n_c$ open.

\begin{figure}[h]
\vspace*{-0mm}
\centerline{\psfig{file=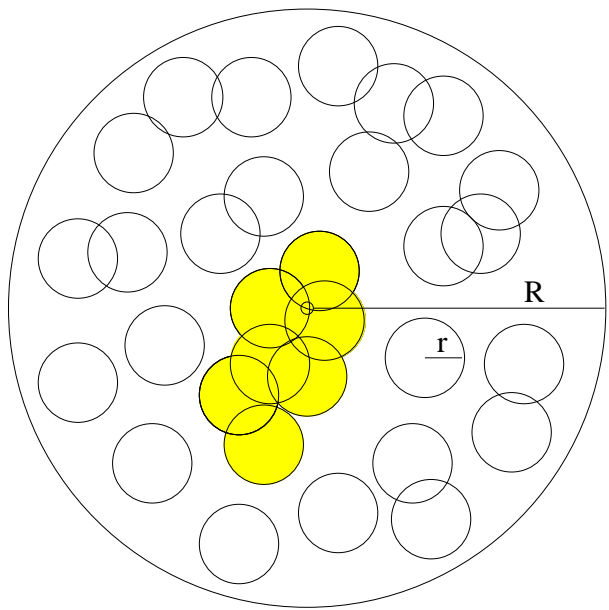,height= 80mm,angle= -90}}
\caption{Clustering of discs in a circular area.}
\end{figure}

\medskip

Consider now a central high energy collision of two identical heavy
nuclei, during which essentially all nucleons undergo several
interactions. These occur in such rapid succession that the nucleons
cannot `recover' between successive interactions. Each individual
nucleon-nucleon collision establishes a colour flux tube or string
between the collision partners; the additive quark model
\cite{quarkmodel} suggests that at present (SPS) energies
it connects two triplet colour charges. The transverse radius $r$ of
such strings is expected to have the form \cite{Luescher,Laermann}
\be
r^2 = {1 \over \pi \sigma} \ln (L/L_0), \label{3}
\ee
where $L$ is the separation between the colour charges and $L_0 ~\lsim~
0.1 - 0.3$ fm the `string formation' length \cite{Alvarez,Schilling};
$\sigma \simeq 0.16$ GeV$^2$ denotes the string tension. Extensive
lattice QCD studies find that in the range $0.5 < L <2$ fm, $r \simeq
0.2 - 0.3$ fm and varies at most weakly \cite{Laermann,Schilling}. In a
dense environment, colour screening will moreover constrain the possible
values of $L$, so that here a radius of the mentioned size appears
appropriate. For nuclear collisions in their early stage, we thus obtain
a spaghetti-like structure of intertwined QCD strings, and a
cut in the transverse plane results in the picture shown in Fig.\ 1,
giving a distribution of transverse string areas over a circle of
nuclear area. We want to study what happens when these overlap more and
more to form connected spatial regions of increasing size in the
transverse plane\footnote{The first study of this kind was carried out
for a specific phenomenological string model \cite{Pajares}.}; hence the
appropriate framework is the two-dimensional case of percolation theory
introduced above.

\medskip

We shall now first study how the cluster density in the transverse plane
varies as function of the density of discs, then how the geometric
cluster size is related to the cluster density. It is useful to define
the dimensionless cluster density
\be
\n = \left({N_{\rm cl} \over \a}\right) a  \geq 1, \label{4}
\ee
where $N_{\rm cl}$ denotes the average number of discs making up the
cluster of size \A; by definition, $\n = 1$ for an isolated disc.
Similarly, we define dimensionless cluster size measures
\be
\s = {\a \over a } \geq 1 \label{5}
\ee
and
\be
\S = {\a \over A} \leq 1; \label{6}
\ee
these are normalized such that $\s=1$ for an isolated
disc and $\S = 1$ when the cluster covers the whole basis area $A$.
To simplify notation, we shall from now on always mean the
average over many configurations when we speak of quantities such as
density, cluster density or cluster size.

\medskip

We shall see shortly that the cluster density \N~increases smoothly with
the overall density $n$. Since also the cluster size $\s$ grows with
$n$, clusters of a certain density will always have a certain size.
Hence at the onset of deconfinement, at some $\n=\n^c$, clusters will
have an
intrinsic size, $\s^c=\s(\n^c)$. Deconfinement therefore begins in
regions which cannot be arbitrarily small; the new phase makes its first
appearance in an area of finite size. This is the main result of
our note and leads to interesting consequences for the pattern of
\J~suppression by deconfinement. We now turn to cluster formation in
detail.

\medskip

To consider a specific case, we set $R/r  = 20$, corresponding to $R=5$
fm and $r=0.25$ fm. The area of the discs thus is 1/400 of the `nuclear'
cross section $A=\pi R^2$ over which they are distributed; we shall
return later to the role of $R/r $ and its specific value. To determine
average cluster quantities, we calculate the size of the cluster
containing the origin of $A$ (the shaded area in Fig.\ 1) in a given
configuration and then average
over many configurations \cite{Stauffer}. In Fig.\ 2a we see that the
cluster density \N~increases monotonically with the overall density
$n$, starting from unity for $n \to 0$.

\medskip

The crucial result of percolation theory is that the size $\s$ of the
cluster, in contrast to its density, shows a dramatic variation with
$n$. In particular, for $R/r  \to \infty$, $\s$ diverges at the
percolation point $n_p=1.175$, i.e., when the overall density is
somewhat above one disc per disc area; $\S$ vanishes there. Of course
there is a considerable overlap of discs, and at $n_p$, the ratio of
the area covered by discs (in clusters of all sizes) to the overall
area becomes $1-\exp[-1.12] \simeq 0.67$; it converges to unity only
for $n \to \infty$. For finite $R/r $, this critical behaviour will be
modified by finite size effects, and for the value $R/r =20$ chosen
above, we obtain the cluster size behaviour shown in Fig.\ 2b. As
expected, it shows the strongest variation around the percolation point;
for sufficiently large systems it is here governed by Eq.\ (\ref{1}) for $n$
below and Eq.\ (\ref{2}) above $n_p$.

\begin{figure}[h]
\vspace*{-15mm}
\centerline{\psfig{file=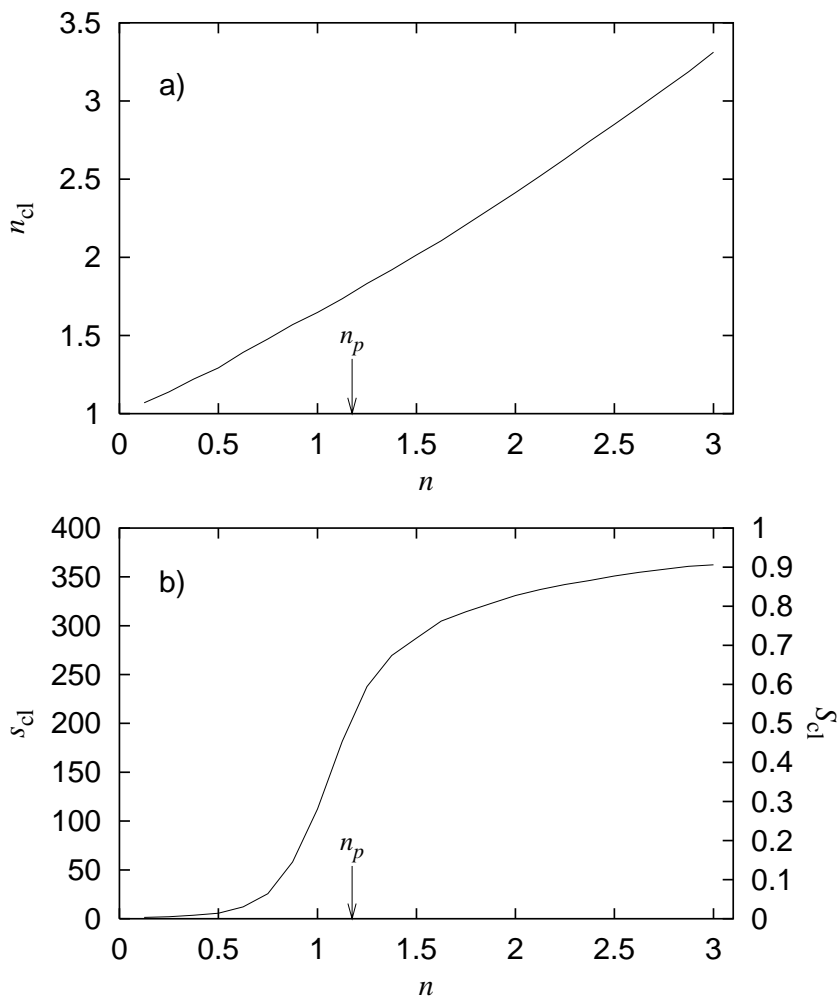,height= 160mm,angle= 0}}
\vspace*{-10mm}
\caption{The cluster density $\n$ (a) and the cluster size $\s(\S)$
(b), as functions of the overall density $n$, for $R$=5 fm, 
$r$=0.25 fm and a uniform random distribution.}
\end{figure}

\medskip

Let us now try to `translate' $n$ and \N~into variables relevant
for deconfinement in nuclear collisions. The number of strings is
determined by the number of nucleon-nucleon collisions; this can be
determined experimentally by measuring the rate of Drell-Yan dilepton
production. Nucleon-nucleon collisions and strings play a basic role in
the early stages of the nucleus-nucleus interaction; the modification of
the parton distributions associated to the onset of deconfinement
can occur through successive nucleon-nucleon interactions, if the time
between the collisions is too short for the restoration of a physical
nucleon \cite{KLNS}. Eventually the strings fuse into `wounded'
nucleons which lead to the observed soft hadron production, and on this
level the number of times a nucleon was hit by others no longer
matters. In a Glauber-based description \cite{KLNS}, both the average
(transverse) collision density and the corresponding density
of wounded nucleons can be calculated for nucleus-nucleus
collisions at fixed impact parameter $b$. Through convolution with the
correlation between the number of wounded nucleons at fixed $b$ and the
energy $E_T$ of the measured secondary hadrons, it then becomes
possible to determine all quantities of interest in terms of
experimental observables. Fig.\ 2a thus effectively shows the collision
density in the average cluster as function of the overall collision
density; the latter can then be correlated to the density of wounded
nucleons and thus to $E_T$.

\medskip

As already noted, deconfinement is expected to set in when there is
enough `internetting' between interacting nucleons \cite{KLNS}, i.e.,
when the collision density within a cluster becomes sufficiently high;
we denote this value by $\n^c$. It is reached at a certain value
$n_c=n(\n^c)$ of the overall density, which can, but does not need
to be the percolation point (see Fig.\ 2a). By Fig.\ 2b, this identifies
the corresponding critical cluster size $\s^c = \s(\n^c)$. In other
words, requiring a critical collision density automatically forces the
cluster to have a certain minimum size. This result
clarifies the question of how much medium is needed before one can
speak about a new phase: systems of geometric size $\s < \s^c$ will on
the average not have a string density sufficient for deconfinement.

\medskip

The onset of deconfinement in nuclear collisions is thus governed by
two parameters: the deconfinement string density $\n^c$ (or some
equivalent quantity) and the transverse string size $r$; with $\n^c$ and
$r$ fixed, we can then determine the size $\s^c$ of the bubbles of
deconfined medium present at the deconfinement threshold. The size of
the nuclear transverse area $\pi R^2$ is given by nuclear
geometry. For a realistic description, we use Wood-Saxon nuclear
distributions \cite{W-S} and a Glauber-based formalism to calculate the
distribution of collisions in the transverse plane as function of the
impact parameter in $A-B$ collisions \cite{KLNS}. The distribution of
the string cross sections in the transverse plane is thus no longer
completely random; it is weighted with the mentioned nucleon-nucleon
collision density.

\medskip

In the remainder of this note, we shall apply our considerations to the
study of \J~suppression \cite{Matsui} in nuclear collisions as observed
in $S-U$ and $Pb-Pb$ data \cite{NA50}. For this
purpose, we will choose $r$ in the range 0.2 - 0.3 fm and fix $\n^c$ such
that central $S-U$ collisions are just below the deconfinement
threshold \cite{B-O}. We can then address two questions:
\begin{itemize}
\item{How is the deconfinement threshold thus defined related to the
percolation point?}
\item{What is the effect of the finite cluster size on the deconfinement
pattern?}
\end{itemize}
As answer to the first question, in previous work \cite{Pajares} the
coincidence of deconfinement and percolation had been assumed,
$n(\n^c)=n_p$. On the second question, a recent study \cite{KNS2} had
already shown that a minimal bubble size for the deconfined medium
leads to an abrupt onset of \J~suppression as function of the collision
centrality. The critical bubble size was there obtained through
superheating in a first order phase transition. Such a picture is,
however, not as easily adapted to the environment produced in nuclear
collisions as the present cluster formation approach, which moreover
does not contain the bubble size as an open parameter.

\medskip

Our application will be based on the successive suppression scenario
\cite{MTM} - \cite{Gupta}, in which the \X~and the \P~are dissociated
essentially at the deconfinement point, the \J~later.
This implies that the fraction
(about 40 \%) of \J's coming from \X~and \P~decay will be suppressed
when $\n=\n^c$, if these states are produced within a deconfined
cluster. The directly produced \J's then survive until $\n^{\rm diss}(\j)$;
based on energy density estimates \cite{MTM,Karsch}, this
is expected to be considerably larger than $\n^c$. Note that in our approach
also the onset of direct \J~suppression will be abrupt, governed by the 
cluster size at $\n^{\rm diss}(\j)$. -- Since the very
loosely bound \P~can in addition also be dissociated by hadronic
comovers \cite{KLNS}, we use below $\n^c$ the experimental suppression
for the 8 \% of the \J~from this state.

\medskip

We now calculate the distribution of nucleon-nucleon collisions and the
corresponding cluster density and cluster size in $S-U$ and $Pb-Pb$
collisions as function of the impact parameter $b$. In the Table, we show
the results for $r=0.20, 0.26$ and 0.30 fm. It is seen that the cluster
density for central ($b=0$) $S-U$ collisions is reached in $Pb-Pb$
collisions for an impact parameter 8.0 fm $<b<$ 8.5 fm. This result is
found to be quite independent of $r$, even though the associated
cluster densities and sizes show a strong $r$-dependence.

\medskip

\begin{center}
\begin{tabular}{|cc|c|c|ccccc|}
\hline
         &           &$r$~(fm)  & S-U  & \multicolumn{5}{c}{Pb-Pb} \vline \\
\hline
 $b$     & (fm)      &          & 0    & 0    &  4   &  8   &  8.5 & 10 \\
\hline \hline
$N_{coll}$&          &          &200   & 910  & 650  &  241 & 200  & 100 \\
$A$      & (fm$^2$)  &          &37.6  & 137.9& 85.6 & 38.8 & 33.5 & 19.3 \\
$N_{coll}/A$ &(fm$^{-2}$)&      &5.32  & 6.60 & 7.59 & 6.21 & 5.97 & 5.17 \\
\hline \hline

$N_{cl}$&            &          & 16.7 &139.1 & 98.5 & 19.9 & 14.8 & 6.6 \\
$A_{cl}$&(fm$^2$)    & 0.20     & 1.47 & 10.9 & 7.81 & 1.70 & 1.29 & 0.62 \\
$n_{cl}/a$&(fm$^{-2}$) &        & 11.2 & 12.6 & 12.4 & 11.3 & 11.1 & 10.2 \\
\hline 
$N_{cl}$&            &          & 91.8 &828.6 &551.1 &110.3 & 78.8 & 21.2 \\
$A_{cl}$&(fm$^2$)    & 0.26     & 11.9 & 91.5 & 62.5 & 14.1 & 10.3 & 3.00 \\
$n_{cl}/a$&(fm$^{-2}$) &        & 7.70 & 9.05 & 8.82 & 7.81 & 7.62 & 6.95 \\
\hline 
$N_{cl}$&            &          & 149.0&878.4 &605.2 &179.9 &136.5 & 40.2 \\
$A_{cl}$&(fm$^2$)    &  0.30    & 23.7 &111.0 & 80.0 & 28.2 & 22.0 & 7.16 \\
$n_{cl}/a$&(fm$^{-2}$) &        & 6.32 & 7.91 & 7.57 & 6.40 & 6.22 & 5.60 \\
\hline

\end{tabular}
\end{center}

\centerline{Table: Collision densities and cluster sizes for $S-U$ and 
$Pb-Pb$ collisions.}

\bigskip

Tuning $r$ in the given range leads to variations in the
size of the cluster at the onset of deconfinement, and this in turn
determines how abruptly the \J~production rate drops at that point. We
find best agreement with the most recent data \cite{NA50R} for $r=0.26$
fm. With a total nuclear overlap area of 35 - 40 fm$^2$ for both
central $S-U$ collisions and $Pb-Pb$ collisions at $b=8$ fm, the cluster
size at the onset of deconfinement is for this $r$ value less than half
the total area; hence the density profile varies little within a given
cluster. The resulting form for the anomalous \J~suppression
is shown in Fig.\ 3.  As direct consequence of the finite cluster
size at the onset of deconfinement we get an abrupt onset of
the suppression as function of impact parameter $b$. As function of the
measured transverse energy $E_T$, the drop is softened by the inherent
$E_T-b$ smearing of the experiment, which is here included in the form
described in \cite{KLNS}. The agreement between our model and the data
is seen to be quite good.

\begin{figure}[h]
%\vspace*{-15mm}
\centerline{\psfig{file=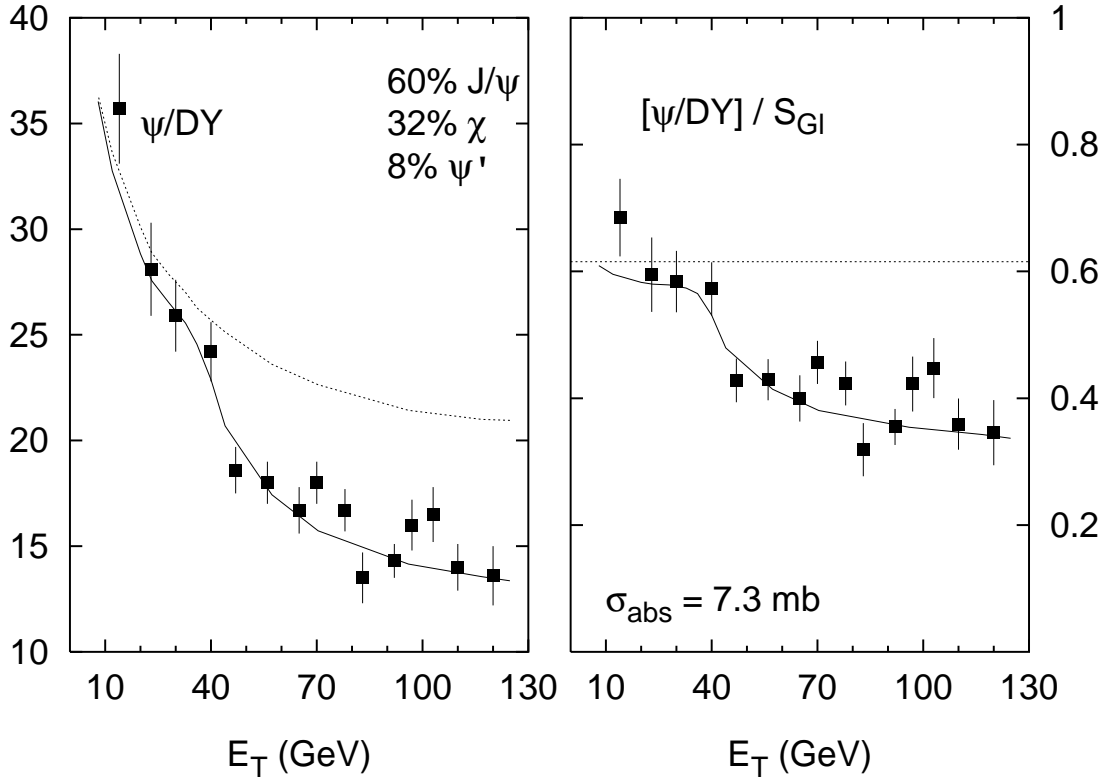,height= 110mm,angle= 0}}
%\vspace*{-10mm}
\caption{The anomalous \J~suppression given by our model, compared to
data \cite{NA50R}; the righthand figure shows the results after division
by pre-resonance absorption \cite{KLNS}.}
\end{figure}

\medskip

Finally we return to the relation of deconfinement and percolation.
For an infinite two-dimensional continuum system, percolation occurs
at $n=1.175$. For $r=0.25$ fm$^2$, we obtain $n_c \simeq 1.2$, for
$r=0.26$ fm$^2$, $n_c \simeq 1.3$; thus the threshold value obtained
from central $S-U$ collisions is quite close to the infinite area
percolation point, supporting the conjecture of percolation as basis 
for deconfinement \cite{Baym}-\cite{Pajares}, \cite{satz}.

\medskip

In conclusion: the onset of deconfinement for a sufficient overlap of
strings appears to provide a conceptually clear picture for the
beginning of quark-gluon plasma formation. Its main and
model-independent consequence is that a critical collision density
implies clusters of a critical size. The resulting abrupt onset
of deconfinement is found to agree with recent data on anomalous
\J~suppression in $Pb-Pb$ collisions.

\bigskip

\centerline{\bf Acknowledgements}

\medskip

It is a pleasure to thank F.\ Karsch, E.\ Laermann and D.\ Stauffer for
helpful comments.

\end{document}